\documentstyle[multicol,pre,aps,epsf]{revtex}
\newcommand{\epsplace}[1]{\epsfxsize=3.25in \epsfbox{#1}}
\newcommand{\be}{\begin{equation}}
\newcommand{\ee}{\end{equation}}
\newcommand{\bea}{\begin{eqnarray}}
\newcommand{\eea}{\end{eqnarray}}

\renewcommand{\caption}[1]{\refstepcounter{figure} \noindent {\bf Figure \thefigure:} {\sf #1}}
\newcommand{\bbmc}{\begin{multicols}{2} }
\newcommand{\eemc}{\end{multicols}}

\title{Coexistence of native and denatured phases in a single protein-like molecule}
\author{Rose Du,${}^{1}$ Alexander Yu. Grosberg,${}^{1,2}$ Toyoichi Tanaka${}^{1}$}

\address{
${}^1$Department of Physics and Center for Materials Science and
Engineering, \\ Massachusetts Institute of Technology, Cambridge,
Massachusetts 02139,  USA \\
${}^2${\em On leave from:\/} Institute of Chemical Physics,
Russian Academy of Sciences, Moscow 117977, Russia
}

\address{ {\em \bigskip \begin{quote}
In order to understand the nuclei which develop during the course 
of protein folding and unfolding, we examine phase segregation 
of a single heteropolymer chain which occurs in equilibrium.
These segregated conformations are characterized by a nucleus of 
monomers which are superimposable upon the native conformation.
We computationally generate the phase segregation by applying a
``folding pressure,'' or adding an energetic bonus for 
native monomer-monomer contacts. 
The computer models reveal 
a fundamental difference in the nucleation process between 
heteropolymeric and the more familiar vapor-liquid systems:
in a polymer system, some % Some 
nuclei hinder folding via topological constraints and 
must be partially destroyed in order for folding to proceed.
To illustrate this finding, we examine the kinetics of protein
unfolding in the long chain limit through scaling arguments.  We find 
that because of the topological constraints, the critical nucleus size 
is of the order of the entire chain size so that unfolding time  
scales as $\exp \left[ c N^{2/3} \right]$, where $N$ and $c$ are the chain length 
and a constant.  
\end{quote} }}

\begin{document}
\maketitle

\bbmc

Proteins fold and unfold cooperatively through the transition accompanied
by a large latent heat and other signs of discontinuity \cite{Privalov}. 
As long as a single protein molecule can be described in statistical terms 
\cite{Lifshitz68},
folding and unfolding should be identified as first order phase
transitions.  According to sophisticated equilibrium statistical
mechanics models \cite{RMP}, 
this sharpness, or cooperativity, of the transition is
a direct manifestation of protein nonrandomness.  Indeed, proteins are
heteropolymers, but with nonrandom sequences that have been
selected, presumably in the course of evolution.  It is now an outstanding
challenge to understand why and how this selection causes proteins to meet
the kinetic requirements of rapid (milliseconds to seconds) and reliable folding.

Since the first order phase transition nature of folding, with its connection 
to the evolutionary selection of sequences, was realized, much attention has  
been paid to nucleation as a natural scenario of folding 
\cite{Fersht_nucleus,Shakhnovich_nucleus,Thirumalai_nucleus}
% (although the traditionally overlooked alternative scenario of 
% spinodal decomposition is also possible).  
The straightforward implementation of the nucleation idea, however appealing, 
faces difficulties, as evidenced by the recent heated debates 
\cite{Wolynes_nucleus_debate,Shakhnovich_nucleus_debate,Thirumalai_nucleus_debate}.   
In hindsight, these difficulties are hardly surprising as
% both heterogeneity and 
chain connectivity must significantly modify the 
very concept of nucleation in ways which we do not yet have the proper 
insight for.

Our goal in the present paper is to gain such an insight by 
reexamining the foundations of the nucleation concept for proteins 
and computationally generating  
conformations containing candidates for nuclei for toy proteinlike 
models.  We will systematically employ the analogy with 
other first order phase transitions, such as the liquid-gas transition, where 
the kinetic concept of nucleation is ultimately related to the 
phase segregated states in equilibrium.  Indeed, the nucleus is  
nothing more than a piece of a ``new'' equilibrium phase which 
tentatively coexists with the surrounding sea of the  ``old'' phase.  
Of course, the nucleus is not really in equilibrium since it grows.  However, and we 
view that as the single major lesson to be learned from a liquid-gas 
type system \cite{L_L_Pitaevskii}, the nucleus is, in the proper sense, 
close to equilibrium.  
Indeed, a nucleus grows slowly - in the sense that all  
other degrees of freedom have sufficient time to relax 
while the nucleus size does not change appreciably.   
The nucleus size can thus be said to be a ``good reaction coordinate.''  In 
terms of landscape theory \cite{Wolynes_landscape}, the relevant profile 
is then that of the free energy taken as a function of the nucleus size.  
The important part of that picture is, again, that the nucleus is 
fairly close to being in equilibrium.     

Thus, we have to address equilibrium phase segregation and phase coexistence 
in a proteinlike molecule.  Note that we are 
referring to the equilibrium coexistence of two distinct phases within a single 
protein chain which should not be confused with the coexistence of 
folded and unfolded molecules in solution.  In order to address the 
phase segregated (macro)states, we resort again to the analogy with 
a liquid-gas system for which there are two
ways to bring the system into the phase segregated state:  one is 
to bring it exactly to the transition temperature, and then add 
(or remove) some heat;  the other involves a range of temperatures 
and the control of pressure and volume.  
The former approach is not usable 
in Monte Carlo simulations since they are done at constant 
temperatures and  
heat transfer cannot be controlled in canonical ensembles.  
As for the latter approach, the analogues of pressure and volume  
have not been defined for proteinlike systems.  This is precisely what 
we will do.   

Recall that the equilibrium folding theory for nonrandom 
(designed \cite{RMP}) heteropolymers recognizes the 
native overlap $Q$ as an order parameter (for any given 
conformation, $Q$ is the number of native monomer-monomer bonds).  
This quantity is thermodynamically additive and is
thus similar to volume in a liquid-gas system.  It is then 
straightforward to define the analogue of pressure, which we call 
folding pressure $P_Q$.  It is the quantity conjugate to $Q$:  
given the Hamiltonian $H_0$ of our protein, 
applying folding pressure $P_Q$ means taking the Hamiltonian 
$H=H_0 -P_Q Q$.  In other words, folding pressure is an additional 
energy bonus for every correct (native) bond.  The $-P_{Q} Q$ term 
can also be thought of as a perturbation
of the original Hamiltonian $H_{0}$ by Go interactions \cite{Go}. 

Although it may not be easy to directly realize pure folding pressure 
in real experiments, changing experimentally controlled 
environmental parameters, such as pH or denaturant concentration, 
affects the folding pressure as well.  In this
sense, examining folding pressure in the theoretical model is 
just as relevant as studying temperature.  Besides, and even more 
importantly, the very simple idea of folding pressure allows us to 
exercise physical intuition in a new fruitful way.  

To perform Monte Carlo simulations of the folding transition,
the polymer is modeled as a self-avoiding chain of 27 or 48  monomers on 
a cubic lattice.  The Hamiltonian is given by
$H_{0}=\sum_{IJ} B_{s_{I}s_{J}} \Delta({\bf r}_{I}-{\bf r}_{J})$
where $I,J$ label  monomers along the chain,
$B_{s_{I}s_{J}}$ is the interaction between species $s_{I}$ and $s_{J}$,
$\Delta({\bf r}_{I}-{\bf r}_{J})=1$ if $I$ and $J$ 
are nearest neighbors
and $\Delta({\bf r}_{I}-{\bf r}_{J})=0$ otherwise.
We employ the model \cite{IIM}
in which the energies $B_{ij}$ are chosen independently from a Gaussian distribution.
The sequence of species along the chain was obtained through
simulated annealing so that the ground state of the polymer
is the native conformation \cite{EugeneDesign,RMP}.

\begin{figure}
\epsplace{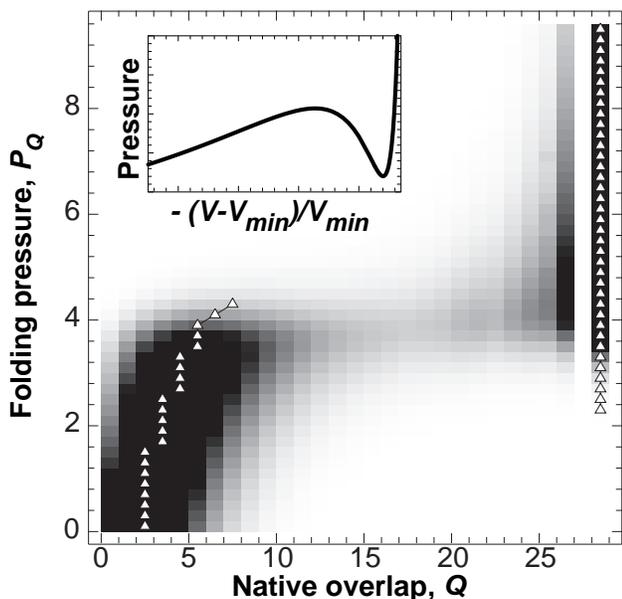}
\caption{
The distribution of a 27-mer over $Q$ at various $P_{Q}$ 
at $T=4.0$.  Both $P_Q$ and $T$ are measured in the 
units of $ \delta \! B$, the variance of the interaction matrix.  
At each $P_Q$, the normalized distribution is given
by the gray level:  darker places correspond 
to higher probability densities.   
% The distribution over $Q$ is bimodal in the region close
% to the transition folding pressure due to metastable states while
% remaining monomodal at low and high pressures.  
Triangles indicate the peaks of the distributions.  The Van der Waals 
isotherm in the inset shows close qualitative similarity.    
\label{fig:pqt40}}
\end{figure}

The $P_{Q}-Q$ isotherms which are obtained by performing
long Monte Carlo runs at various temperatures and folding pressures
are strikingly similar to the $P-V$ isotherms for 
a liquid-gas system (see Figure~\ref{fig:pqt40}).    
The folding pressure appears as the $-P_{Q}Q$ term 
in addition to the usual energy $H_{0}$ in 
the standard Metropolis criteria.  For 27-mers, runs are
at least $2 \times 10^{9}$ (and up to $5 \times 10^{10}$)
iterations,  ie.  until the distribution of conformations
over $Q$ has reached equilibrium.  As we vary the pressure at each temperature,
the distribution of $Q$ changes from a monomodal distribution
to a bimodal distribution near the transition pressure
and back again to a monomodal distribution (Figure~\ref{fig:pqt40}).  The bimodal
distribution is characterized by two maxima at $Q=Q_{max}$ and
$Q=Q_{u}$ and a wide minimum in between.  For 27-mers, $Q_{max}=28$, 
$Q_{u}$ varies from 2 to 8, and the minimum
is centered around $Q=18$. $Q_{max}$ corresponds to 
the folded state and $Q$'s near $Q_{u}$ correspond
to the unfolded states.   
The bimodal distribution occurs 
over a range of pressures, thus manifesting the 
metastable states.

The immediate (trivial) insight following from Figure~\ref{fig:pqt40} 
is that there should be both regimes of nucleation and of spinodal 
decomposition.  Indeed, if the system is initially 
equilibrated, say, in the unfolded phase (lower left corner 
in Figure~\ref{fig:pqt40}) and folding  
is then caused by an instant folding pressure quench at 
constant temperature, the kinetics will proceed differently 
depending on whether the system is quenched to the region 
where the original state is metastable or totally 
unstable.  In particular, we have observed both 
nucleation and spinodal decomposition
by examining the time evolution of the native contacts
close to or far away from the transition temperature, respectively
\cite{Rose_thesis}.
Close to the transition point the polymer remains folded or unfolded for an
extended period of time until a particular group of native contacts which
form the ``nucleus'' or ``critical loop''  are
formed or broken, after which the polymer rapidly folds or 
unfolds, as is characteristic of nucleation.  
Far away from the transition temperature, there is no longer
a free energy barrier to the (un)folded phase during
(un)folding, that is, all domains of the (un)folded
phase are unstable.  This can be seen as a gradual increase
of the (un)folded phase.  The spinodal 
decomposition scenario implies nonspecific 
``homopolymer'' collapse as the first stage of folding  
% Note that homopolymer collapse kinetics is itself a
% complex problem attracting significant attention 
\cite{homopolymer_folding}.

By applying folding pressure, we
can bring the system to the transition point at 
any given temperature.  Once we are at the transition
point, the stage of the transition 
can be controlled by varying $Q$. This is analogous to 
controlling the volume in a liquid-gas system by moving the piston 
when the pressure is equal to the transition pressure.
This naive ``piston'' model is particularly useful due 
to the finite size of the system.  Indeed, for macroscopic 
systems, the critical nucleus size is always much smaller 
than the system size.  Accordingly, an infinitesimally small 
change in volume or displacement of
a piston, would be needed to produce an equilibrium phase segregated 
state in which the size of, say, the liquid phase would be 
approximately 
that of a critical nucleus.  Since the polymers 
of interest are of moderate size, the expected critical nucleus 
is not negligibly small compared to the entire system.  This is 
why the interesting stages of the transition are seen when $Q$ is 
significantly different from the values of any of the phases.  

We computationally generated the phase segregated 
\eemc
\begin{figure}
%\epsplace{cloud.eps}
\epsfxsize=\textwidth \centerline{\epsffile{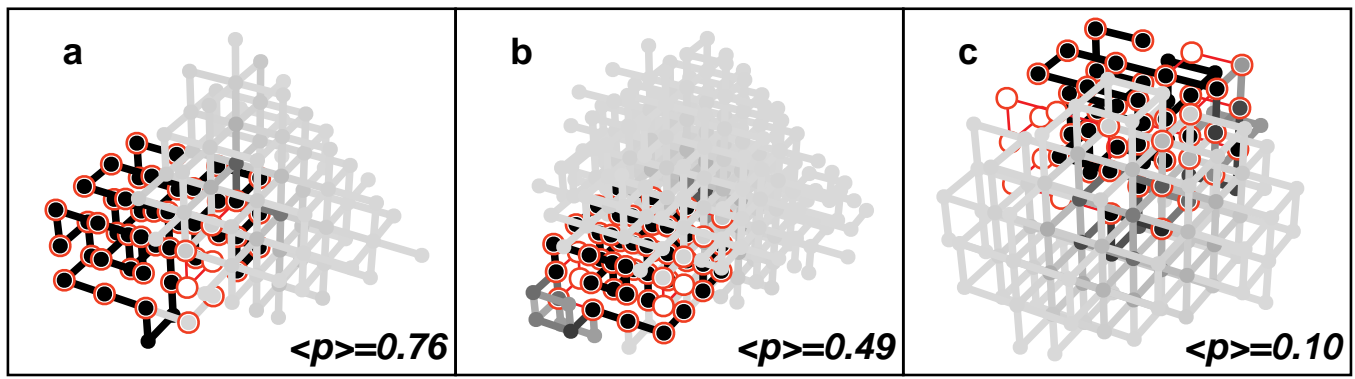}}
\caption{
Phase segregated macrostates. Empty circles outline the native conformation. 
Gray level for every position shows how frequently this position is occupied by 
a monomer, black means that a monomer is always there.    
% Black indicates that a monomer always occupies that position 
% while white indicates that a monomer rarely occupies 
% that position.  
Folding probabilities, $p$, for microstates 
belonging to each of these three macrostates have fairly narrow 
distributions:  average values of $\left< p \right>$ are shown in each 
figure, while the variance is significantly smaller, about 
$\left< \Delta p \right> \simeq 0.05$ for all three examples.  
\label{fig:cloud}
}
\end{figure}
\bbmc
\noindent 
macrostates of a heteropolymer  
by running simulations at the transition point 
($T=1.7$, $P_{Q}=0.5$ for 27-mers and $T=2.28$, 
$P_{Q}=1.0$ for 48-mers; both $T$ and $P_Q$ are given in units of 
$\delta \! B$, the variance of the interaction matrix \cite{IIM}).  
To mimic a fixed ``piston'' position, we restrict the  
value of $Q$ to some chosen level ${\overline Q}$.  
We first obtain conformations at 
$Q={\overline Q}$ by doing a Monte Carlo run with unconstrained $Q$ and 
collecting conformations every time the system passes through the ${\overline Q}$ 
``surface.''   
Each of the collected conformations 
with $Q={\overline Q}$  can be used to initiate 
a new restricted run in which 
$Q={\overline Q}$.  The set of microstates 
(individual conformations) 
encountered in the restricted run will 
form a macrostate at the transition
temperature $T$ characterized 
by the given values of $P_Q$, and $Q={\overline Q}$.  
% Three examples of such 
% macrostates are shown in Figure \ref{fig:cloud}.  
The phase segregation is clearly seen in Figure \ref{fig:cloud} where 
% each microstate is superimposed on the native conformation
% so as to obtain the 
orientation and position are chosen for each microstate with maximum superposition
with the native state.  
When the microstates are subsequently superimposed upon one another,
we see that there is a set of ``core'' monomers, mostly superimposed
upon the native state, which do not
fluctuate in position (Figure~\ref{fig:cloud}).  
These monomers can be interpreted
as the ``native'' phase, while the monomers which fluctuate
belong to the ``denatured'' phase.  
% Thus we have macrostates each of which is composed of two coexisting
% phases within a single heteropolymer chain.          

Based on the analogy with the liquid-gas system, we expect that phase 
segregated macrostates obtained at subsequent 
values of ${\overline Q}$ 
(``piston positions'') would be very similar to those encountered 
subsequently in time during a real kinetic event.  To test this 
expectation, we employ the general method suggested in the work 
\cite{Reaction_Coord} and measured the folding probability 
$p$ for each of the microstates belonging to the phase segregated
macrostate.  We found that the distribution of $p$'s is indeed 
very narrow for most of the macrostates.  In the 
three examples shown in Figure~\ref{fig:cloud}, the variance 
of $p$ is as low as $\left< \Delta p \right> \simeq 0.05$.  
This is to be compared with the  
very broad distributions of $p$'s for the ensemble of all states
with the given $Q = {\overline Q}$, where $\left< \Delta p \right>$ 
can be of order unity \cite{Reaction_Coord}.  This means 
that by ``moving the piston'' in the way described above, we indeed 
drag the system along its natural kinetic path.  However, as we 
mentioned, this is only valid for a majority of phase segregated 
macrostates.  There are important exclusions from the rule
which represent the fundamental difference between 
(hetero)polymeric system and the more familiar liquid-gas one.  

\begin{figure}
\epsplace{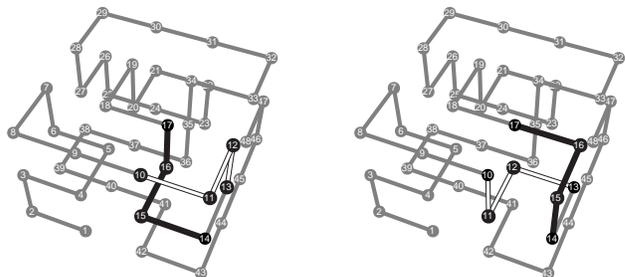}
\caption{
Example of topologically hindered folding.   The left figure is the 
native conformation.  The right figure represents a 
% conformation 
% which has been taken from a series of 
phase segregated states obtained  
by gradualy increasing the proportion of the folded part.  While 
% as many as 
41 out of 47 polymeric bonds (87\%) are successfully 
superimposed on the native conformation, 
% so that the system 
% can be said to be $41/47 = 87\%$ native, 
the denatured phase developed 
a topological constraint. 
% which hinders folding.  
Folding 
% from this conformation
can proceed only if the nucleus (folded phase) is partially destroyed.  
\label{fig:constraint}}
\end{figure}

To understand the problem, consider Figure~\ref{fig:constraint}.  It 
demonstrates one of the examples in which ``pushing the piston''
brings the system to a dead end  
instead of dragging it along its path from one phase to the 
other.  The reason for the deadlock is 
purely topological: while the amount of the folded phase is growing 
at the expense of the denatured phase, a topological constraint 
has been formed in the latter.  As a result, the system arrives at 
a conformation which cannot fold without first destroying a 
significant part of its correctly folded ``native phase.''

\begin{figure}
\epsplace{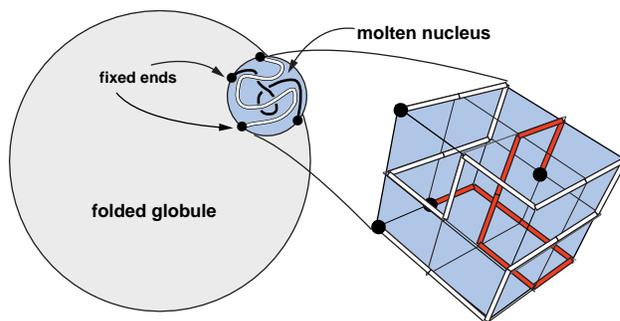}
\caption{
Schematic representation of unfolding nucleus inside a large folded 
globule 
and its lattice model.  
% The right figure explains the use of the $3 \times 3 \times 3$ 
% lattice model to simulate two-loops configurations with frozen ends.  
\label{fig:topogran}}
\end{figure}

The insight we gained from considering phase segregated macrostates 
and the role of topological constraints can be now used to 
produce a scaling analysis for very long ($N \to \infty$) 
chains, thus approaching the problem formulated 
in the work \cite{scaling_of_time}.  Here, we restrict 
ourselves to the scaling of {\em unfolding} time, which is 
more instructive in terms of the role of topological 
constraints.  We begin with a correctly folded 
native globule, and then quench the temperature and/or folding 
pressure such that native state becomes (globally) unstable.  
The nucleus of the unfolded denatured phase, which sooner or later 
will appear in this system can be imagined as a ``bubble" 
% (either inside globule or close to its surface) 
of polymer melt (or solution; Figure~\ref{fig:topogran}).  
This polymer liquid consists 
of one or several loops.
% of the original polymer.  
It is important to note that on the time scale 
of interest, while the nucleus remains unchanged in size, the ends 
of all those loops are firmly quenched at the corresponding 
``root points'' on the surface of native phase which is frozen.

Since all ends are fixed, the topology of the  
loops is well-defined and quenched.  
The crucial point here is not the classification of the given state as ``entangled'' or
``unentangled''
but that the mutual positions of the loops are quenched in the same 
topological class to which it belonged in the native state 
(unless a chain end belongs to the nucleus, which is improbable 
in the $N \to \infty$ limit).  This fact has profound consequences 
in terms of the critical nucleus size and, accordingly, the unfolding 
barrier height.  Indeed, normally, in a vapor-liquid system, 
nucleation free energy
can be schematically written as $- \alpha \Delta \! T N + \sigma N^{\frac{2}{3}}$, 
where the volume part ($N$), which is proportional to the 
deviation from the transition point $\Delta \! T$ (degree 
of overheating or overcooling in the initial quench) is negative 
(favorable) and the surface part ($N^{\frac{2}{3}}$) is positive (unfavorable).  
For a polymer,  melting of the nucleus  does not release 
all of the volume free energy  $- \alpha \Delta \! T N$
because the melted part remains topologically constrained.  Thus, there 
appears an additional positive contribution to the 
nucleus free energy.  As long as the nucleus remains small 
compared to the entire globule, this new term can be estimated 
as $+T\ln {\cal M}$, where ${\cal M}$ is the number of topologically 
different classes. 
% for the given number of loops and loop lengths.  
Since ${\cal M}$ grows exponentially with the number of monomers 
involved, 
we end up with an extra volume term which is always positive and 
independent of $\Delta \! T$: 
$- \alpha \Delta \! T N + \beta T N + \sigma N^{\frac{2}{3}}$.  To estimate 
$\beta$, we performed an exhaustive enumeration of all two-loops 
conformations with fixed ends within a $3 \times 3 \times 3$ cube 
($N=27$, see Figure~\ref{fig:topogran}) and found that $\beta = 0.45$ is  
of order unity and by no means small (contrary to an early estimate 
\cite{Finkelstein}).  
Thus, topological constraints significantly increase the height of the 
barrier, in complete agreement with our simulations 
(Figure~\ref{fig:constraint}).  As long as 
% overheating, or 
temperature 
jumps causing unfolding, $\Delta \! T$, remain finite and $N$-independent, 
the critical nucleus is of the order of the entire globule, 
and thus the unfolding time scales as 
$\exp \left[ c \cdot N^{2/3} \right]$, as opposed to some $N$-independent 
time for the phantom polymer, which is allowed to freely pass through 
itself.  As for  $c$, it is a constant, and we see no grounds to assume 
that it is significantly different from unity.  Our result 
agrees very well with the original estimate of folding time under equilibrium 
conditions given by Finkelstein, but sharply 
contradicts to its latest improvement \cite{Finkelstein}.         

To conclude, we found that study of the phase segregation occuring 
in an equilibrium proteinlike heteropolymer sheds light on the 
possible and impossible nuclei configurations relevant for folding 
and unfolding kinetics.   We found in particular that topological 
constraints play an important role in determining the critical 
nucleus.  In the case of unfolding, topological constraints 
dramaticaly increase the size of the critical nucleus, causing the 
unfolding time to scale exponentially with the chain length.  

The work was supported by NSF grant DMR-9616791.

\eemc
\end{document}